\documentclass[twocolumn,superscriptaddress,floatfix,
longbibliography,aps,pra,preprintnumbers]{revtex4-1}

\usepackage{graphicx}
\usepackage{bm}
\usepackage{xcolor}
\usepackage{epstopdf}
\usepackage{amsmath}
\usepackage{amssymb}
\usepackage{epstopdf}
\usepackage[normalem]{ulem} 
\usepackage{enumerate}

\usepackage[urlcolor=blue,colorlinks=true,citecolor=blue,linkcolor=blue,pdfstartview={FitH},bookmarks=false]{hyperref}

\definecolor{darkgreen}{HTML}{008000}

\newcommand{\Tr}{\text{Tr}}

\graphicspath{{fig/}{./fig/}{.}}

\begin{document}

\title{Probing the chirality of 1D Majorana edge states around \\ a 2D nanoflake in a superconductor}

\author{Andrzej Ptok}
\email[e-mail: ]{aptok@mmj.pl}
\affiliation{\mbox{Institute of Nuclear Physics, Polish Academy of Sciences,
ul. W. E. Radzikowskiego 152, PL-31342 Krak\'{o}w, Poland}}

\author{David J. Alspaugh}
\email[e-mail: ]{dalspa1@lsu.edu}
\affiliation{Department of Physics and Astronomy, Louisiana State University, 
Baton Rouge, Louisiana 70803-4001, USA}

\author{Szczepan G\l{}odzik}
\affiliation{\mbox{Institute of Physics, Maria Curie-Sk\l{}odowska University,
Plac Marii Sk\l{}odowskiej-Curie 1, PL-20031 Lublin, Poland}}

\author{Aksel Kobia\l{}ka}
\affiliation{\mbox{Institute of Physics, Maria Curie-Sk\l{}odowska University,
Plac Marii Sk\l{}odowskiej-Curie 1, PL-20031 Lublin, Poland}}

\author{Andrzej M. Ole\'{s}}
\affiliation{\mbox{Institute of Theoretical Physics, Jagiellonian University,
Profesora Stanis\l{}awa \L{}ojasiewicza 11, PL-30348 Krak\'ow, Poland}}
\affiliation{Max Planck Institute for Solid State Research,
Heisenbergstrasse 1, D-70569 Stuttgart, Germany}

\author{Pascal Simon}
\email[e-mail: ]{pascal.simon@u-psud.fr}
\affiliation{Universit\'e Paris-Saclay, CNRS, Laboratoire de Physique des Solides, F-91405, Orsay, France}

\author{Przemys\l{}aw Piekarz$\,$}
\email[e-mail: ]{piekarz@wolf.ifj.edu.pl}
\affiliation{\mbox{Institute of Nuclear Physics, Polish Academy of Sciences,
ul. W. E. Radzikowskiego 152, PL-31342 Krak\'{o}w, Poland}}

\date{\today}

\begin{abstract}
The interplay between superconductivity, magnetic field and spin--orbit 
coupling can lead to the realization of non--trivial topological phases.
Recent experiments 
have found signatures of such phases in magnetic nanoflakes formed by nanostructures coupled to a superconducting substrate.
These heterostructures comprise  a topologically non-trivial region  surrounded by a trivial one due to the finite magnetic exchange field induced by the magnetic nanoflake. 
The analysis of the topological phase diagram of such a system shows that  a similar phase separation occurs by tuning the chemical potential of the nanoflake.
In this paper, we study such a possibility in detail, analyzing the spatial extent of the edge modes circulating around the nanoflake and discussing some practical implementations. 
We also show how the chirality of Majorana edge states can be probed using scanning tunneling spectroscopy with a double tip setup.
\end{abstract}

\maketitle

\section{Introduction}
The quest for the realization of Majorana zero
modes (MZMs), driven by the pursuit of both fundamental physics and their potential application to fault-tolerant topological quantum computation~\cite{kitaev.03,nayak.08,alicea.11,alicea.12,aasen.hell.16}, is steering an active research in engineering $p$-wave superconductivity.
Non-Abelian braiding is an essential step towards topological
quantum computing, though it has not yet been experimentally achieved with MZMs. 
Due to the localized nature of MZMs, their braiding will necessarily involve both coupling and manipulation processes.

However, it has been suggested that non-Abelian braiding is not only restricted to MZMs, but can also be implemented  with one-dimensional (1D) chiral Majorana fermions~\cite{lian.18}.
Chiral Majorana fermions can manifest themselves as quasiparticle
edge states of a two-dimensional (2D) topological $p$-wave superconductor \cite{qi.zhang.11,alicea.12}. 
Signatures of 1D chiral Majorana quasiparticles have been recently observed in 2D heterostructures consisting of a quantum anomalous Hall insulator bar in contact with a superconductor~\cite{he.17}.
Additionally, recent progress in atomic scale engineering~\cite{menard.guissart.17,drost.ojanen.17,kim.palaciomorales.18,kamlapure.corils.18,menard.mesaros.18}  is opening up new perspectives for the practical implementation of chiral Majorana fermions by spatially building  non--trivial topological phases separated from trivial ones.
Recent progress includes Co islands grown on a Si substrate and covered by a monolayer  of Pb~\cite{menard.guissart.17} and nano-scale Fe islands of  monoatomic height on a Re surface~\cite{palaciomorales.mascot.18}.
Due to the non--trivial topological phase transition resulting from a gap closure, in-gap edge states surrounding the topological superconducting (SC) domain are observed. 
In both experiments, these in-gap states are strongly delocalized around the islands and have been interpreted as signatures of chiral Majorana fermions.

\begin{figure*}[!t]
\centering
\includegraphics[width=0.825\linewidth]{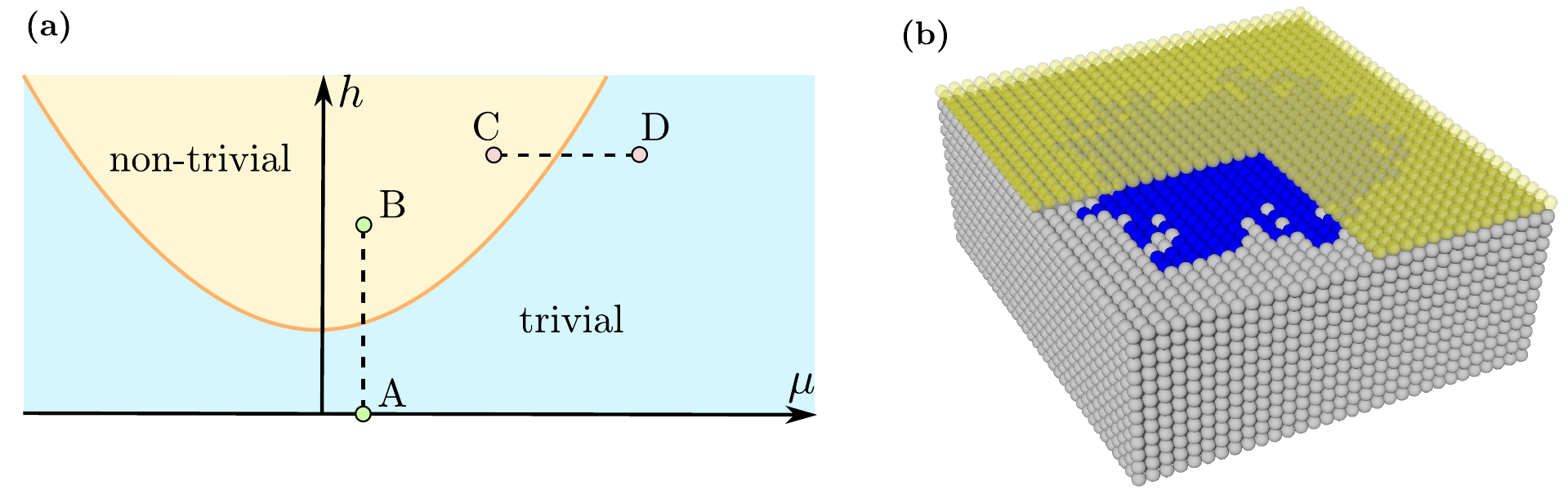}
\caption{
(a) Schematic representation of the topological phase diagram as a 
function of the chemical potential $\mu$ and a magnetic Zeeman energy $h$.
(b) Schematic view of the studied system. A superconducting (SC) layer
(yellow transparent atoms) is deposited on a magnetic substrate (gray atoms). 
The nanoflake is formed by non-magnetic atoms located at random sites either 
between the substrate and the SC layer as shown in the figure or above the 
SC layer. For a constant SC gap $\Delta$, the boundary between the 
topologically trivial (blue region) and non--trivial phase (yellow region) 
is given by a parabola $h_c^2=\Delta^2+\mu^2$ [orange line in (a)].
\label{fig.schemat}
}
\end{figure*}

In 2D superconductors with Rashba spin--orbit coupling (SOC), the transition to a non--trivial phase can be induced by an external Zeeman magnetic field \mbox{\cite{sato.fujimoto.09,sato.takahashi.09,sato.takahashi.10}.} 
The boundary between the trivial and non--trivial topological phases is given by $h_c^2=\mu^2+\Delta^2$,
see Fig.~\ref{fig.schemat}(a), where $h_c$ stands for the critical Zeeman field for given values of the doping $\mu$ and the SC gap $\Delta$. 
In the aforementioned experimental results, the Zeeman magnetic energy arises from the presence of magnetic dopants interacting with the substrate, while the SOC and the SC gap are intrinsic to the subsystem. 
Looking at the phase diagram presented in Fig.~\ref{fig.schemat}(a), one observes that a line which connects the points A and B in the ($\mu,h$) plane could correspond to an inhomogeneous system in real space, where a non-magnetic trivial domain (point A) surrounds or borders with a topological magnetic domain (point B).

In this paper, we instead choose to explore  an alternative route. 
We consider an inhomogeneous system but in a constant magnetic Zeeman energy, which would reside on the C--D line in Fig.~\ref{fig.schemat}(a).
The transition to the topological domain occurs due to a change in the chemical potential $\mu$.
Such a system could be constructed experimentally in many different ways: one option would be to substitute the magnetic atoms in the Co island~\cite{menard.guissart.17} by non-magnetic ones and add a magnetic field parallel to the SC Pb monolayer [Fig.~\ref{fig.schemat}(b)].
Another promising approach is to use the versatility offered by 2D van-der-Waals heterostructures~\cite{Novoselov.16}.
A possible way to generate an homogeneous Zeeman exchange energy in the proximity of a superconductor could be engineered by stacking recently synthesized 2D magnetic materials~\cite{Burch.18,Gong.19} with a transition metal dichalcogenide~(TMD) superconductor such as NbSe$_2$.
A non-magnetic island can be obtained by evaporating some alkaline adatoms to enforce charge transfer.

The paper is organized as follows:
In Sec.~\ref{sec.circ_geo}, we first start with a circular geometry for the nanoflake and derive the dispersive chiral Majorana edge states analytically in the continuum limit.
We also discuss the spatial extent of the chiral Majorana modes in the tranverse direction.
In Sec.~\ref{sec.tbm_res} we compare our results obtained in the continuum limit to exact diagonalization of a tight binding model on a lattice. 
In Sec.~\ref{sec.exp_chiral}, we propose and study a setup in order to measure the chirality of the Majorana edge states using two ferromagnetic tips.
Finally, we present a summary of our results and give conclusions in Sec.~\ref{sec.sum}.


\section{Nanoflake with circular geometry}
\label{sec.circ_geo}

\begin{figure}[!b]
\centering
\includegraphics[width=\linewidth]{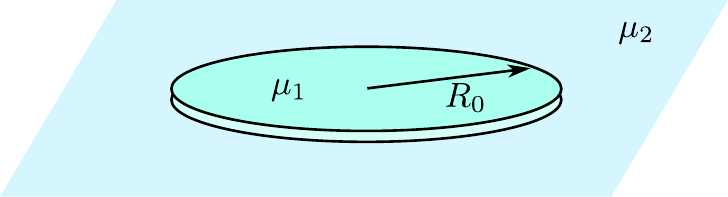}
\caption{
Schematic representation of the discussed system in the thermodynamic 
limit with a circular nanoflake deposited on a substrate.
\label{fig.cylinder}
}
\end{figure}

We begin by considering a nanoflake with a circular symmetry as depicted 
in Fig.~\ref{fig.cylinder}. For simplification and without a loss 
of generality, we can also assume a smooth boundary of the nanoflake. 
Keeping in mind the circular symmetry, the Hamiltonian will commute with 
the $z$th component of the total angular momentum operator $J_z\equiv L_z+S_z$. 
We may then find the energies of the bound state wave functions localized 
at the edge of the nanoflake in terms of the $m_{J}$ quantum numbers. 
This method allows us to determine the existence of chiral subgap states 
within our system, and has successfully been used in the studies of other 
2D systems with circular symmetry such as graphene \cite{asmar.ulloa.14}.

Thus, the real-space normal state Hamiltonian has the form,
\begin{eqnarray}
\nonumber H(\bm{r},\bm{\nabla})=
\left(-\dfrac{\nabla^{2}}{2m}-\mu(\bm{r})\right)\sigma_{0} 
+\alpha(\bm{\sigma}\times -i\bm{\nabla})_z + h\sigma_z , \\
\label{eq.ham_1st}
\end{eqnarray}
where $\sigma_{i}$ (for $i=\{0,x,y,z\}$) are the Pauli matrices acting 
in spin space. Here, the system is 2D with ${\bf r}\equiv(x,y)$, and 
the chemical potential is given by,
\begin{eqnarray}
\mu(r) = 
\begin{cases} 
\mu_{1} & r < R_{0} , \\
\mu_{2} & r \geq R_{0} .
\end{cases}
\end{eqnarray}
We may also define the discontinuity $\delta\mu=\mu_{2}-\mu_{1}$ at the boundary,
see Fig. \ref{fig.cylinder}. 
In other words, $\delta\mu$ corresponds to the spatial variation of the chemical 
potential induced by the nanoflake. The BdG Hamiltonian may then be expressed as,
\begin{eqnarray}
\nonumber \mathcal{H} = \dfrac{1}{2}\int d\bm{r}\,\Psi^{\dagger}(\bm{r})
\begin{pmatrix}
H(\bm{r},\bm{\nabla}) & i\sigma_{y}\Delta
\\ -i\sigma_{y}\Delta & -H^{T}(\bm{r},-\bm{\nabla})
\end{pmatrix} 
\Psi(\bm{r}) , \\
\end{eqnarray}
where $\Psi(\bm{r})\equiv (\psi_{\uparrow}(\bm{r}),\psi_{\downarrow}(\bm{r}),
\psi_{\uparrow}^{\dagger}(\bm{r}),\psi_{\downarrow}^{\dagger}(\bm{r}))^T$ 
is the Nambu spinor, with $\psi_{\sigma}(\bm{r})$ being electron field 
operators which destroy an electron with spin $\sigma$ at location ${\bm r}$.

Due to the circular symmetry of the nanoflake (or more precisely, the 
scalar chemical potential $\mu(\bm{r})=\mu(r)$), the BdG Hamiltonian commutes 
with the $z$th component of the total angular momentum operator $J_z=L_z+S_z$. 
It follows that the Hamiltonian and $J_z$ share the same eigenstates. 
The eigenstates of $J_z$, with the half-integer eigenvalues $m_J$, are given by
\begin{eqnarray}
\varphi_{m_{J}} = 
\begin{pmatrix}
u_{m_{J}\uparrow}(r)  \,e^{i(m_{J} - 1/2)\theta} \\
u_{m_{J}\downarrow}(r)\,e^{i(m_{J} + 1/2)\theta} \\
v_{m_{J}\uparrow}(r)  \,e^{i(m_{J} + 1/2)\theta} \\
v_{m_{J}\downarrow}(r)\,e^{i(m_{J} - 1/2)\theta}
\end{pmatrix}.
\end{eqnarray}
To focus on states with small total angular momenta, we take a 
low-energy approximation and neglect the kinetic energy term in the 
Hamiltonian~\cite{oreg.refael.10}. By writing the $u_{m_{J}\sigma}(r)$ 
and $v_{m_{J}\sigma}(r)$ functions in terms of the Modified Bessel 
Functions of the First and Second Kind, we may then solve for the bound state 
wave functions localized at $r=R_0$. The details of this approach can be found 
in the Supplemental Material (SM)~\footnote{See Supplemental Material at [URL will be 
inserted by the publisher] for details of the analytical calculations and additional 
numerical results for the nanoflake with irregular shape. In the second case, 
the numerical results for different sets of values $\lambda$ and $V_0$ are presented.}, 
and the resulting wave functions have the form:
\begin{eqnarray}
\varphi_{m_{J}} = 
\begin{cases} 
\varphi_{m_{J}1} & r < R_{0} , \\
\varphi_{m_{J}2} & r \geq R_{0} ,
\end{cases}
\end{eqnarray}
where
\begin{widetext}
\begin{eqnarray}
\varphi_{m_{J}1} &=& \sum_{\eta=\pm}N_{m_{J}1\eta} \begin{pmatrix}
a_{m_{J}1\eta} I_{m_{J} - \tfrac{1}{2}}(k_{m_{J}1\eta}r) e^{i(m_{J} - \tfrac{1}{2})\theta} \\
b_{m_{J}1\eta} I_{m_{J} + \tfrac{1}{2}}(k_{m_{J}1\eta}r) e^{i(m_{J} + \tfrac{1}{2})\theta} \\
c_{m_{J}1\eta} I_{m_{J} + \tfrac{1}{2}}(k_{m_{J}1\eta}r) e^{i(m_{J} + \tfrac{1}{2})\theta} \\
I_{m_{J} - \tfrac{1}{2}}(k_{m_{J}1\eta}r) e^{i(m_{J} - \tfrac{1}{2})\theta}
\end{pmatrix},
\end{eqnarray}
and
\begin{eqnarray}
\varphi_{m_{J}2} &=& \sum_{\eta=\pm}N_{m_{J}2\eta} \begin{pmatrix}
a_{m_{J}2\eta} e^{i(m_{J}-\tfrac{1}{2})\pi}  K_{m_{J} - \tfrac{1}{2}}(k_{m_{J}2\eta}r) e^{i(m_{J} - \tfrac{1}{2})\theta} \\
b_{m_{J}2\eta} e^{i(m_{J}+\tfrac{1}{2})\pi}  K_{m_{J} + \tfrac{1}{2}}(k_{m_{J}2\eta}r) e^{i(m_{J} + \tfrac{1}{2})\theta} \\
c_{m_{J}2\eta} e^{i(m_{J}+\tfrac{1}{2})\pi}  K_{m_{J} + \tfrac{1}{2}}(k_{m_{J}2\eta}r) e^{i(m_{J} + \tfrac{1}{2})\theta} \\
e^{i(m_{J}-\tfrac{1}{2})\pi}  K_{m_{J} - \tfrac{1}{2}}(k_{m_{J}2\eta}r) e^{i(m_{J} - \tfrac{1}{2})\theta}
\end{pmatrix}.
\end{eqnarray}
Here $I(z)$ and $K(z)$ are the Modified Bessel Functions of the first and 
second kind, respectively. The $a$, $b$, and $c$ parameters along with the 
normalizations are derived within the SM~\cite{Note1}, while the radial 
momenta which control the spatial extent of the bound states wave functions are given by
\begin{eqnarray}
k_{m_{J}j\eta} &=& \dfrac{1}{\alpha} \sqrt{ h^{2} - E_{m_{J}}^{2} + \Delta^{2} - \mu_{j}^{2} + 2\eta\sqrt{E_{m_{J}}^{2}\mu_{j}^{2} + \Delta^{2}(h - \mu_{j})(h + \mu_{j})} },
\end{eqnarray}
\end{widetext}
where $j = 1,2$ and $\eta=\pm$. For each of the $\varphi_{m_{J}}$ bound states 
localized at $r = R_{0}$, the total radial spatial extent $\xi_{m_J}$ of the wave 
functions are thus determined by 
\begin{equation}
\xi_{m_J}=\max\{k_{m_{J}1+}^{-1},k_{m_{J}1-}^{-1},k_{m_{J}2+}^{-1},k_{m_{J}2-}^{-1}\}.
\end{equation}

\begin{figure}[!t]
\centering
\includegraphics[width=\linewidth]{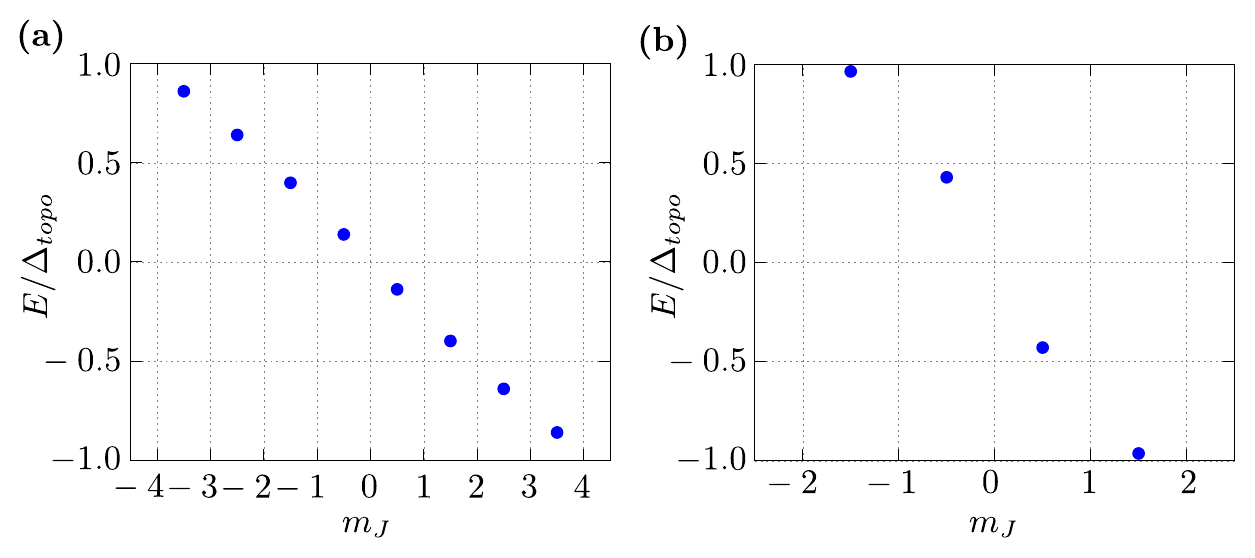}
\caption{
Spectrum of the in-gap dispersive states as a function of the total angular 
momentum quantum number $m_{J}$. \\ Results are presented for two different 
sets of parameters: \\
(a) $\mu_1=0.2$~meV, $\mu_2=0.4$~meV, $\Delta=0.3$~meV, and \\
(b) $\mu_1=0.1$~meV, $\mu_2=0.2$~meV, $\Delta=0.2$~meV.\\
We take $R_0=10$~nm and $\alpha=0.25$~meV$\cdot$nm.
\label{fig.mj}
}
\end{figure}

\begin{figure}[!t]
\centering
\includegraphics[width=\linewidth]{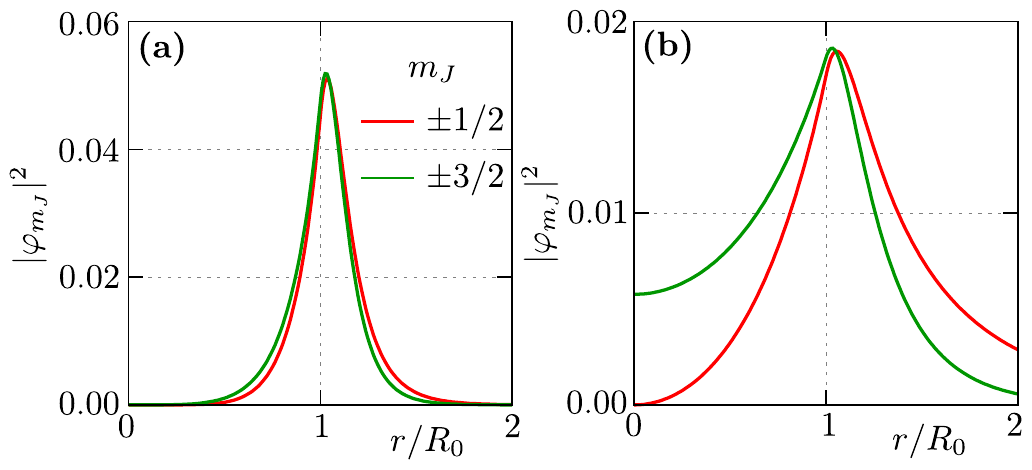}
\caption{
Localization of the in-gap bound states around the edge of the nanoflake for 
different values of the total angular momentum $m_{J}$. Figures (a) and (b) 
correspond to the same sets of parameter values as detailed in Fig.~\ref{fig.mj}. 
\label{fig.local_mode}
}
\end{figure}

The energy spectrum of these bound states vs the $m_{J}$ quantum numbers is 
presented in Fig.~\ref{fig.mj}, while the spatial profile of the wave functions 
is given in Fig.~\ref{fig.local_mode} for two different characteristic sets of 
parameters. The number of in-gap states is quantized due to the finite perimeter of the nanoflake and their energy spacing depends on intrinsic parameters. In the first set of parameters, we find 8 in-gap states while we 
have 4 in-gap states in the second set.
We choose the strength of the Zeeman field such that the topological gap, 
\begin{equation}
\Delta_{\rm top}=\left| h - \sqrt{\Delta^2+\mu^2} \right|,
\end{equation}
is equal both inside and outside the nanoflake. From our first set of data 
we obtain $\Delta_{\rm top}=0.070$ meV, while for our second set we obtain 
$\Delta_{\rm top}=0.0296$ meV. The existence of only one branch of states 
in the topological gap signals that they are chiral (here left handed). 
This in turn supports the hypothesis of a phase separation in space with a chiral Majorana edge state circulating around the nanoflake.

We also note that these dispersive states are not perfectly linear, but instead 
exhibit a slight cubic behavior. Comparing both set of parameters, we see 
that the larger the discontinuity $|\delta\mu|=|\mu_1-\mu_2|$, the more 
localized the edge states are. This behavior is expected; indeed, the in-gap 
Majorana states are pinned at the domain wall which is controlled by the 
spatial variation of the chemical potential.


\section{Tight binding formulation}
\label{sec.tbm_res}

In order to confirm our calculations performed in the continuum limit and to go 
beyond the circular symmetry assumption for a nanoflake, we have also performed
numerical calculations on a lattice based on a tight binding description.
Our system can be described by the following tight binding Hamiltonian,
\begin{eqnarray}
\label{H}
\mathcal{H}=\mathcal{H}_{\rm kin}+\mathcal{H}_{\rm SO}
+\mathcal{H}_{\rm prox}+\mathcal{H}_{\rm flake}.
\end{eqnarray}
The first term corresponds to a free particle on a 2D square lattice,
\begin{eqnarray}
\mathcal{H}_{\rm kin} = \sum_{ij\sigma}\,\left[-t+(4t-\mu-\sigma h)\delta_{ij}\right]\,c_{i\sigma}^{\dagger}c_{j\sigma}^{}.
\end{eqnarray}
Here $t$ is the hopping integral between nearest neighbor sites
\footnote{In the case of an homogeneous 2D system, this gives a 
dispersion relation $\mathcal{E}_{\bm k}=-2t(\cos(k_x)+\cos(k_y))$. 
Then, the minimal available energy is equal to $-4t$.}, 
$\mu$ is the chemical potential (calculated from the bottom of the band), 
while $h$ can be regarded either as a genuine Zeeman energy or as a magnetic 
exchange energy depending on the situation under consideration. 
In all cases, we treat it as an effective magnetic field in what follows.
The second term describes the in-plane SOC,
\begin{eqnarray}
\mathcal{H}_{\rm SO}= 
-i\alpha\sum_{ij\sigma\sigma'} c_{i+{\bm d}_j\sigma}^{\dagger}\left[(
{\bm d}_j\times\hat{\bm\sigma})\cdot\hat{z}\right]_{\sigma\sigma'}^{}
c_{i\sigma'}^{},
\end{eqnarray}
where vectors ${\bm d}_i\in\{\pm\hat{x},\pm\hat{y}\}$ stand for the 
locations of the neighbors of the {\it i}-th site, while 
$\hat{\bm\sigma}=(\sigma_x,\sigma_y,\sigma_z)$ is the vector with the 
Pauli matrices being its components. A SC gap can be induced in the 
layer through the proximity effect---this process is described through 
the third term by the BCS-like form,
\begin{eqnarray}
\mathcal{H}_{\rm prox}=\Delta\sum_i\left(
c_{i\uparrow}^{\dagger}c_{i\downarrow}^{\dagger}+{\rm H.c.}\right).
\end{eqnarray}

The last term in Eq. \eqref{H} denotes the influence of the nanoflake 
at the particle distribution,
\begin{eqnarray}
\mathcal{H}_{flake}=-\sum_i V_i^{} c_{i\sigma}^{\dagger} c_{i\sigma}^{}.
\end{eqnarray}
We assume that every atom comprising the nanoflake changes the energy levels 
of the rest of the sites, i.e., we assume a long-range impurity potential 
given by $V_i=V_0\sum_m\exp\left(-R_{mi}/\lambda\right)$, where the summation 
is carried out over all adatoms in a given configuration 
$\mathcal{V}$~\cite{maska.sledz.07}. Here, 
$\lambda$ denotes the characteristic length of decay of the impurity potential.

The Hamiltonian $\mathcal{H}$ can be diagonalized by the unitary transformation,
\begin{eqnarray}
c_{i\sigma}=\sum_n \left(u_{in\sigma}^{}\gamma_n^{} 
-\sigma v_{in\sigma}^{\ast}\gamma_n^{\dagger}\right) ,
\end{eqnarray}
which leads to the Bogoliubov--de~Gennes (BdG) equations~\cite{bdg} 
of the form
\begin{equation}
\mathcal{E}_{n} \Phi_{in}=\sum_j\mathbb{H}_{ij}\Phi_{jn}, 
\end{equation}
with eigenvectors $\Phi_{in}=\left(u_{in\uparrow},v_{in\downarrow},
u_{in\downarrow},v_{in\uparrow}\right)^T$. Here
\begin{eqnarray}
\mathbb{H}_{ij} = \left( \begin{array}{cccc}
H_{ij\uparrow} & D_{ij} & S_{ij}^{\uparrow\downarrow} & 0 \\
D_{ij}^{\ast} & -H_{ij\downarrow}^{\ast} & 0 & S_{ij}^{\downarrow\uparrow} \\
S_{ij}^{\downarrow\uparrow} & 0 & H_{ij\downarrow} & D_{ij} \\
0 & S_{ij}^{\uparrow\downarrow} & D_{ij}^{\ast} & -H_{ij\uparrow}^{\ast}
\end{array} \right)
\end{eqnarray}
is the Hamiltonian in the matrix form, with matrix elements 
$H_{ij\sigma}= -t\sum_j\delta_{\langle i,j\rangle} 
+ ( 4t -\mu -\sigma h -V_i)\delta_{ij}$ as the kinetic term, 
$D_{ij}=\Delta\delta_{ij}$ describing the SC correlations, and 
$S_{ij}^{\sigma\sigma'}=-i\alpha\sum_{j}\left[ 
({\bm d}_j\times\hat{\bm\sigma})\cdot\hat{z}\right]_{\sigma\sigma'} 
\delta_{\langle i,j\rangle}$ standing for the matrix representation 
of the spin--orbit coupling. 
More details of this method can be found e.g. in Ref.~\cite{balatsky.vekhter.06}.

\subsection{Numerical results}
\label{sec.num_res}

We report the calculations on a $N_x\times N_y=59\times 59$ square 
lattice with periodic boundary conditions (PBC). Omitting generality, 
in the calculations presented in this subsection we use a nanoflake 
with a circular shape and a radius $R_{0}=15.1$, which covers about 
20\% of the total area. We present results for a nanoflake characterized 
by $\lambda=1$ and $V_0/t=-0.06$. In subsequent calculations, we take 
$\alpha/t=0.15$, $\Delta/t=0.3$, and $\mu/t=0.4$. 
If not mentioned explicitly in the text, the value of the magnetic Zeeman 
field $h$ was chosen so that the boundary between the trivial and non--trivial 
phases remains in the center of the artificial domain wall (typically $h/t\simeq 0.4$).

\begin{figure}[!b]
\centering
\includegraphics[width=\linewidth]{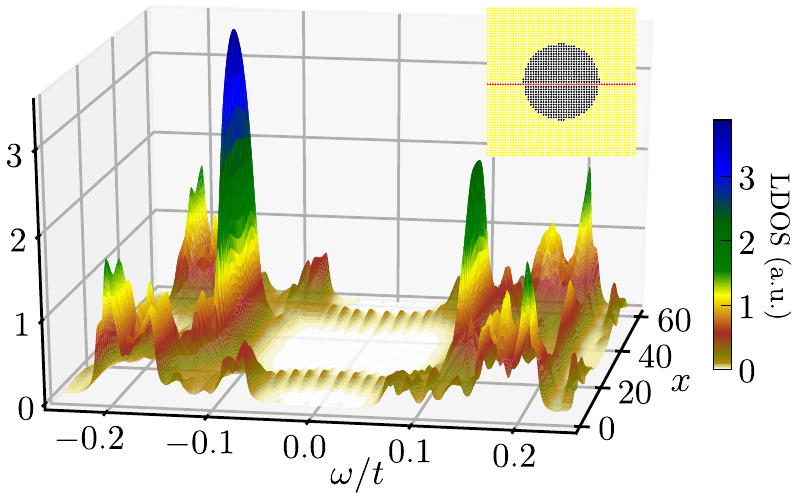}
\caption{
Local density of states along the line presented in the inset.
\label{fig.ldos_c}
}
\end{figure}

\subsubsection{Density of states}
\label{sec.dos}

\begin{figure}[!b]
\centering
\includegraphics[width=\linewidth]{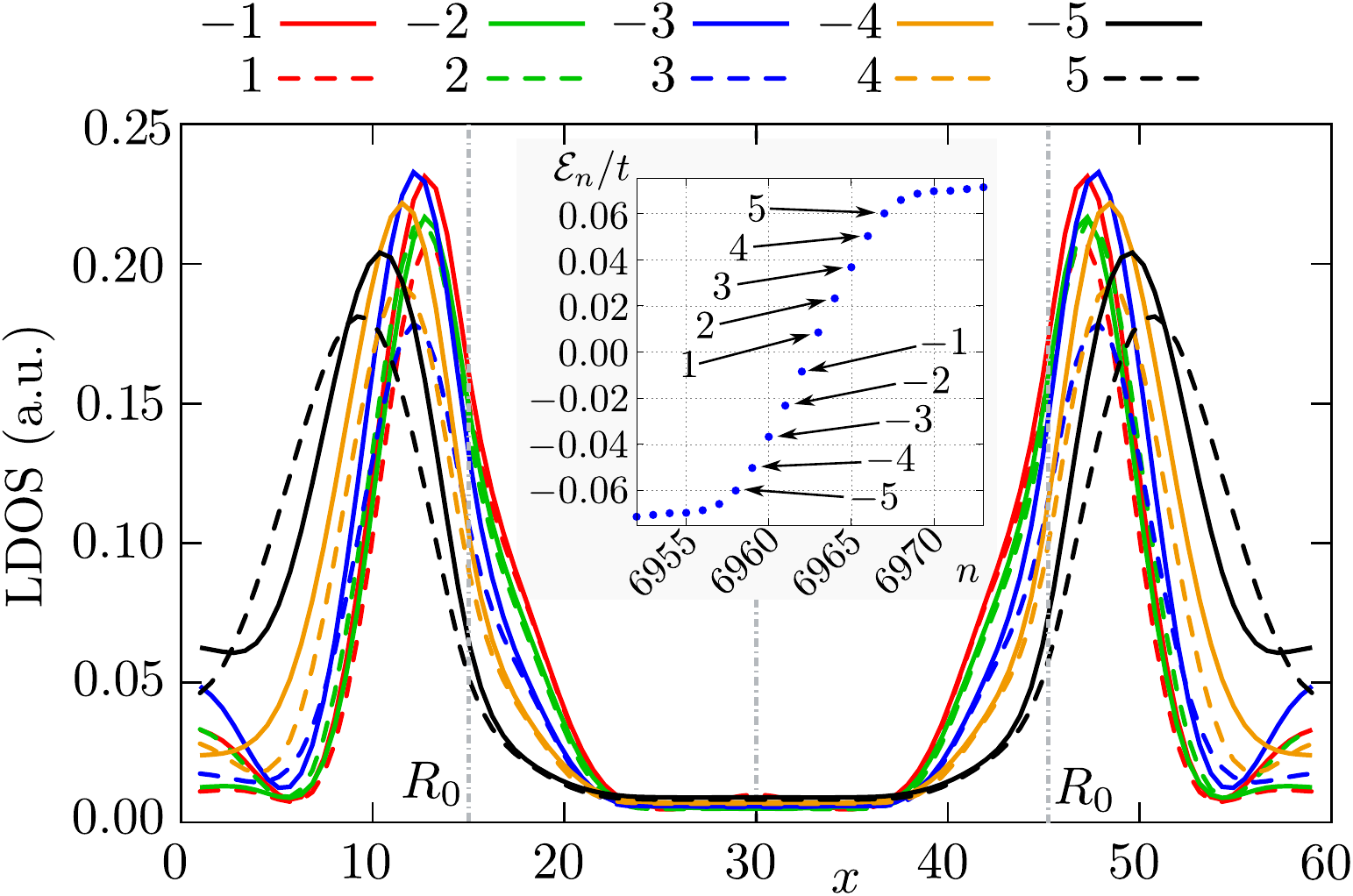}
\caption{
LDOS of several in-gap eigenstates along center of the nanoflake ($y=30$).
The radius $R_{0} = 15.1$ is shown by a gray line (cf. Fig.~\ref{fig.local_mode}).
The inset shows a spectrum of the system and a description of the eigenstates
(cf. Fig.~\ref{fig.mj}).
Numbers from $-5$ to $5$ enumerate the ten states near the Fermi level.
\label{fig.local_mode_tb}
}
\end{figure}

From the solutions of the BdG equations, we first calculate the local density of 
states (LDOS)~\cite{matsui.sato.03},
\begin{eqnarray}
\rho_{i}(\omega)\!&=& 
\sum_{\sigma n}\left[|u_{in\sigma}|^2\delta\left(\omega-\mathcal{E}_n\right) 
+|v_{in\sigma}|^2\delta\left(\omega+\mathcal{E}_n\right)\right]\!, \quad
\label{eq.ldos}
\end{eqnarray}
where we replace the Dirac function $\delta(\omega)$ by a Lorentzian, 
$\delta(\omega)\!=\zeta/[\pi(\omega^2+\zeta^2)]$, 
with a small broadening \mbox{$\zeta\!=0.003t$.}
Figure \ref{fig.ldos_c} shows an example of the LDOS for a chosen path 
(along the nanoflake with $y=30$). The non--trivial domain is separated 
from the trivial phase by in-gap states strongly localized along the 
edge of the nanoflake (cf. Fig.~\ref{fig.local_mode_tb}). 
The domain wall is visible in the form of two sets of LDOS peaks with oscillating intensity near the $\omega/t \simeq 0$, which are a result of the discrete nature of the in-gap state's spectrum.
The LDOS in our system does not exhibit an ``X''--shape crossing through 
the energy gap around the nanoflake edge, in contrast to the results 
presented in Refs.~\cite{menard.guissart.17,bjornson.blackschaffer.18}.
Nevertheless, our results are in agreement with experimental results 
presented in Ref.~\cite{palaciomorales.mascot.18}. In consequence, we 
do not observe the two-ring localization like in Refs.~~\cite{menard.guissart.17,bjornson.blackschaffer.18} but a 
one-ring structure along the nanoflake boundary like in 
Ref.~\cite{palaciomorales.mascot.18}.

The localization of the in-gap states is shown Fig.~\ref{fig.local_mode_tb}.
As we can see, the edge states are localized around the nanoflake at the border between the trivial and topological phases.
All states create localized states with a circular shape of the same radius (equal approximately to $17$), while the radius of the nanoflake is equal $R_{0} \simeq 15.1$ (shown by doted-dashed line).
The difference of these results with respect  to the ones discussed in the continuum limit is a consequence of the smearing of the domain wall given by $V_i$. 

\begin{figure}[!b]
\centering
\includegraphics[width=\linewidth]{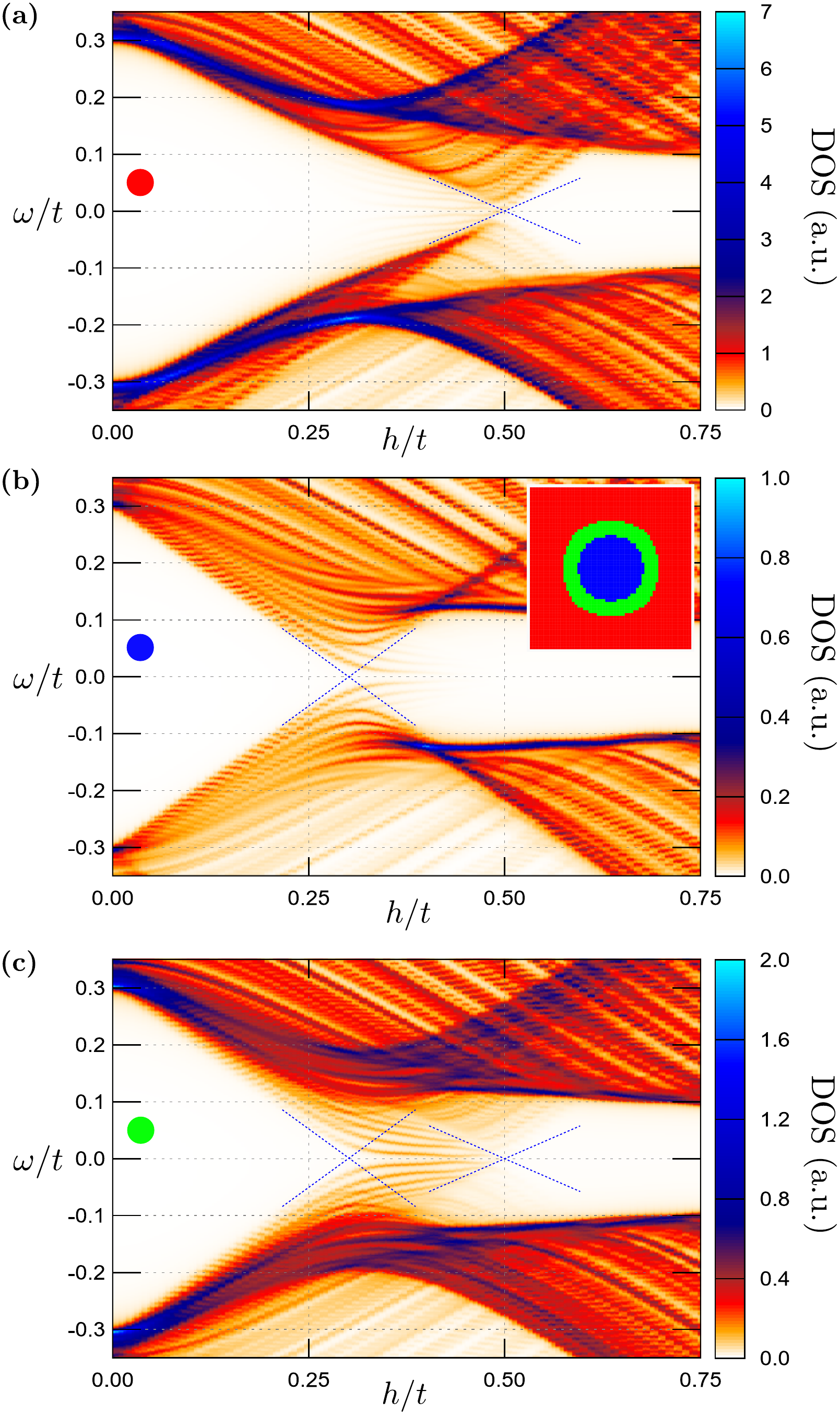}
\caption{
Partial density of states for different regions: 
(a) bulk, 
(b) nanoflake, and 
(c) domain wall. 
Blue dashed lines serve as a guide to the eye and present the linear 
continuation of the gap closing.
\label{fig.pdos_c}
}
\end{figure}

The total density of states (DOS) of the system per site is given by a
summation of the LDOS over the whole 2D space, i.e., 
$\rho(\omega)= 1/N \sum_i\rho_i(\omega)$.
Here, we can separate the sum into three different terms
\begin{eqnarray}
\sum_i\rightarrow
\sum_{i\in\text{Nanoflake}}+\sum_{i\in\text{DW}}+\sum_{i\in\text{Bulk}},
\end{eqnarray}
a contribution from the sites belonging to nanoflake, a contribution from the domain wall (DW), and a contribution from the 
bulk states  [cf. inset in Fig.~\ref{fig.pdos_c}(b)]. 
As in Ref.~\cite{bjornson.blackschaffer.15}, we can define the functions $\mathcal{C}_i$
in order to classify the states in real space.
These functions are equal to $1$ when a site $i$ belongs to a given region, and $0$ otherwise.
We assume that the NF (bulk) region is located in sites where the nanoflake 
changes (does not change) the chemical potential significantly, i.e.,
$| V_i | < 0.95 \max\{| V_i |\}$ ($| V_i | > 0.05 \max\{| V_i |\}$).
Otherwise, we treat the site as a part of the DW region, i.e., when 
$0.95\max\{|V_i|\}\leq |V_i|\leq 0.05\max\{|V_i|\}$. 
These conditions can be smoothed arbitrarily without changing the results qualitatively (see Fig.~S6 and Fig.~S7 in the SM~\cite{Note1}).

We use this recipe to present the partial density of states (PDOS), 
which is defined as follows
\begin{equation}
\tilde{\rho}(\omega)=\sum_i\mathcal{C}_i\,\rho_i(\omega).
\end{equation}
The contribution of the bulk (NF) presented in Fig.~\ref{fig.pdos_c}(a) [Fig.~\ref{fig.pdos_c}(b)], looks like the familiar DOS of a pure 2D 
Rashba spin-orbit coupled superconductor, in which the gap closing is 
followed by a reopening of the topological gap. Dashed blue lines 
serve as a guide to the eye, showing the linear continuation of the 
last negative and first positive eigenvalue.
The value of the magnetic field $h$ for which these dashed lines cross  zero 
energy indicate the phase transition from trivial 
to non--trivial phase.

For effective fields larger than this crossing point, in both cases we 
observe a topological gap opening around 
$\omega/t\simeq\pm 0.1$. Contrary to this, the DW PDOS  
[Fig.~\ref{fig.pdos_c}(c)] ressembles a non--trivial 2D system with egdes 
\cite{kobialka.domanski.19}. 
Between critical fields for bulk and NF 
regions (dashed blue lines crossing the Fermi level), we observe the 
in-gap states associated only with DW, which confirms the existence of
bound states localized along the DW. Additionally, with the increase of 
$h$, the contribution of these states to the total DOS is shifted from the DW 
to bulk region (cf. Fig.~S8 in the SM~\cite{Note1}).

\begin{figure}[!b]
\centering
\includegraphics[width=\linewidth]{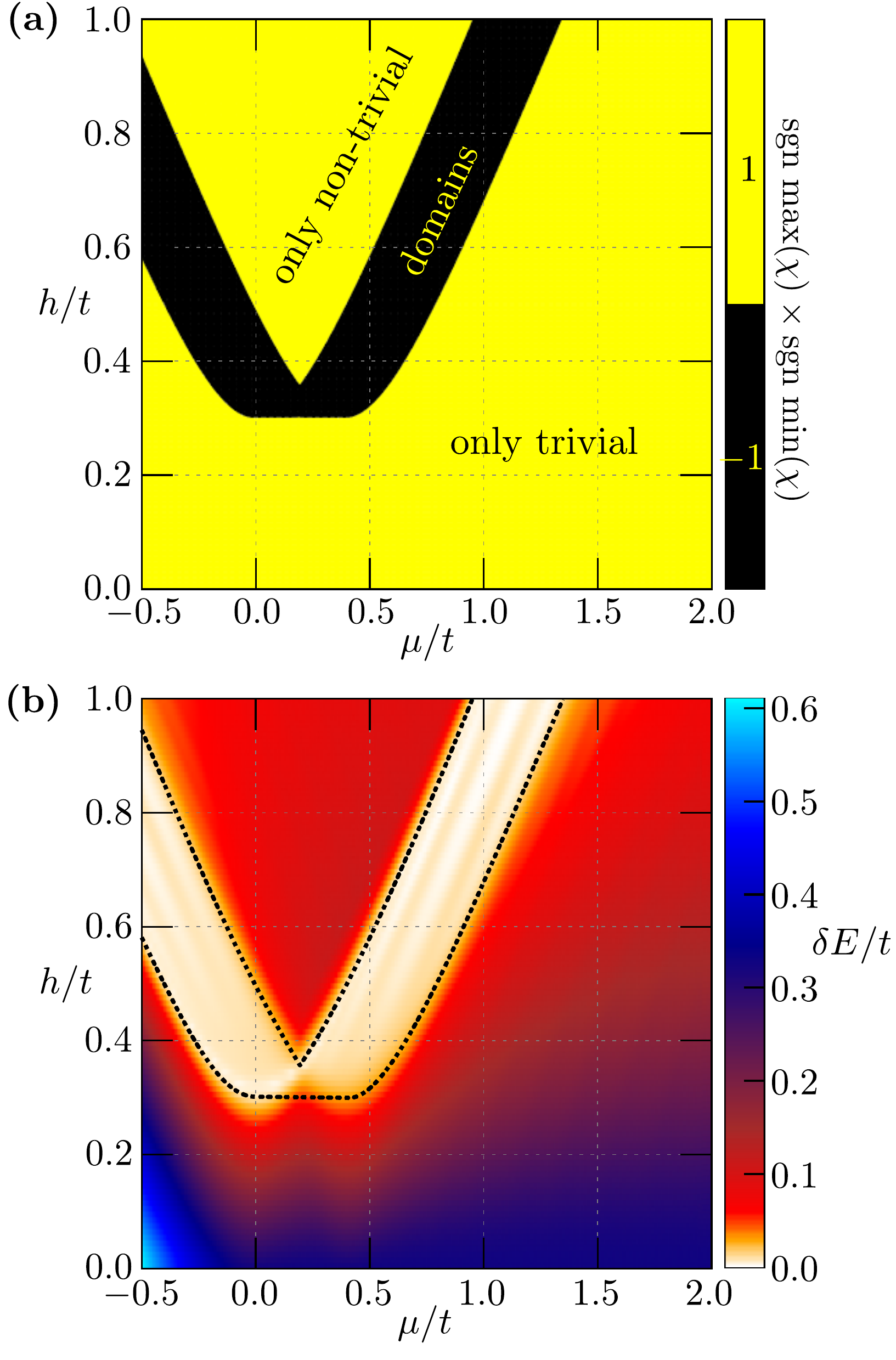}
\caption{
Topological phase diagram--- in the $(\mu,h)$ plane: 
(a) as obtained from the indicator $\chi_i$ defined in Eq. \eqref{eq.ind} and 
(b) the value of the energy gap $\delta E$.
Note that the values in the bottom left corner are very high and exceed 
the scale of the plot.
\label{fig.topo_pd_c}
}
\end{figure}

\subsubsection{Non--trivial topological domains and topological phase diagram}
\label{sec.topo_phase_diag}

A magnetic field, $h$, leads to a closing 
of the trivial SC gap and reopening of a new non--trivial topological 
gap. This occurs at the critical energy $h_c^2=\mu^2+\Delta^2$
~\cite{sato.fujimoto.09,sato.takahashi.09,sato.takahashi.10} 
($h_c/t\simeq 0.5$ for our choice of parameters). In our system, the 
value of the chemical potential varies from site to site i.e., 
$\mu_i=\mu+V_i$. This non-homogeneity can lead to a situation in which 
the above condition is met only locally 
\cite{liu.hu.12,liu.drummond.12,ptok.cichy.18}. We can therefore 
construct a space dependent indicator~\cite{zhou.mohanta.19} that 
describes the spatial distribution of the non--trivial topological 
phase,
\begin{eqnarray}
\label{eq.ind} 
\chi_{i}=\sqrt{ ( \mu + V_{i} ) ^{2} + \Delta^{2} } - h.
\renewcommand{\thefigure}{\arabic{figure}S} 
\end{eqnarray}
Thus, positive (negative) sign of $\chi_i$ indicates the topologically 
trivial (non--trivial) phase.
Indeed, from the analysis of $\chi_i$ under an increase of the  magnetic field $h$ (see Fig.~S9 in the SM~\cite{Note1}), we  find that the non-trivial phase exists in the system when $\chi_i < 0$.

From the above analysis, it follows that the spatial dependence of the 
$\chi_i$ indicator gives a correct information about the emergence of the 
non--trivial topological domain inside the nanoflake. 
Using this condition, we  construct a topological phase diagram in the two-parameter space defined by the chemical potential 
$\mu$ and the effective magnetic field $h$ shown in
Fig.~\ref{fig.topo_pd_c}(a).

First, we recall that the trivial and non--trivial topological phases are 
separated by a parabolic boundary \mbox{$h_c^2=\mu^2+\Delta^2$} in the 
homogeneous system. In our system, the nanoflake introduces a 
non-homogeneity in $\mu$ to the system, thus the boundary splits due to the 
existence of two regions in space where 
topological phases can 
emerge. As a consequence, the boundary in the phase diagram evolves 
into a stripe, whose width is given by $\max |V_i|$. Thus, in the limit 
$V_0\rightarrow 0$ (homogeneous system without any nanoflake), the ``stripe'' would narrow 
down into a line $h_c(\mu)$, as seen in Fig.~\ref{fig.schemat}.

Second, the region of the stripe separating the phases coincides with 
the value of the gap $\delta E$, calculated as the energy difference between 
the eigenvalues closest to Fermi level [Fig.~\ref{fig.topo_pd_c}(b)].
The existence of a non--trivial domain in the system is a result of 
the occurrence of in-gap states with exponentially small eigenenergies 
(described by $\delta E$).

\begin{figure}[!b]
\centering
\includegraphics[width=\linewidth]{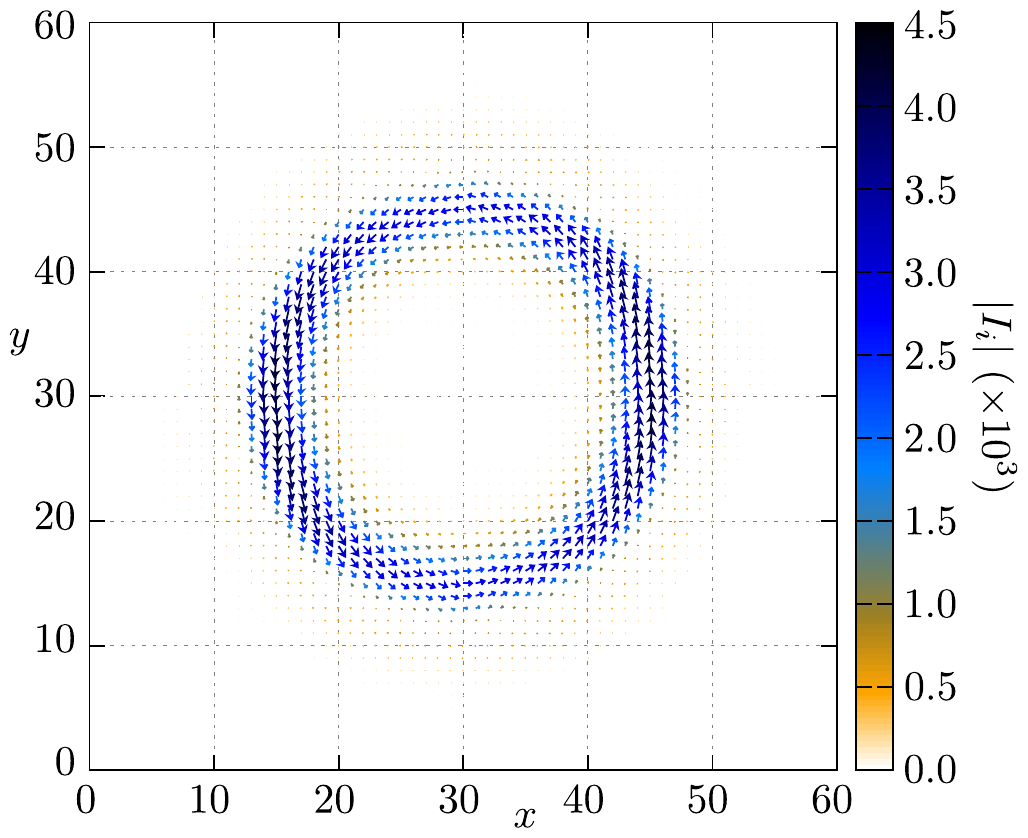}
\caption{
Vector map of the bond current in the described system.
Color corresponds to the absolute value of the current~$| I_i |$
(right scale).
\label{fig.current_c}
}
\end{figure}

\subsubsection{Bond current}
\label{sec.bond_curr}

When the system has boundaries 
\cite{rontynen.ojanen.15,kobialka.domanski.19}, or if artificial 
barriers are introduced~\cite{ptok.cichy.18,kobialka.ptok.19}, in-gap states localize on the edges of the system. The in-gap states 
are localized in a collection of preferable locations, independently of the broadening of domain wall.
These well--localized in-gap states provide a contribution to the bond current  which can be expressed as the local charge flow and  is obtained from the Heisenberg equation 
\cite{pershoguba.bjornson.15,li.neupert.16,bjornson.pershoguba.15},
\begin{eqnarray}
\imath \hbar \frac{\partial \langle n_{i} \rangle}{\partial t} = 
\left\langle [ \mathcal{H},n_{i} ] \right\rangle.
\end{eqnarray}
The current vector field can be represented as a sum of the 
spin-dependent currents, $I_i=\sum_{i\sigma} I_{i\sigma}$, where 
\mbox{$I_{i\sigma}=\partial_t\langle n_{i\sigma}\rangle$} can be 
expressed by the BdG 
eigenvectors~\cite{bjornson.pershoguba.15,glodzik.domanski.18}.

In Fig.~\ref{fig.current_c}, we present the real space map of the bond  
current. Color of the arrows denotes their magnitude $\propto |I_i|$.
Once again, the width of the domain wall (controlled by $\lambda$) is 
reflected by an observable, however, this time it is through the area 
in which there is a significant flow of charge 
(cf. Fig.~S3 in the SM~\cite{Note1}). 
These low energy modes bear a close resemblance to the surface states 
of three-dimensional (3D) topological insulators (TIs)
\cite{hasan.kane.10,chang.li.16}. 
Due to the SOC induced band inversion and 
bulk-boundary correspondence, the 2D surface of a TI hosts metallic 
states which disperse through the band gap.
Generally, band inversion 
is a result of the non--trivial phase transition, but here
``surface states'' are limited to the boundary of the nanoflake, which 
serves as an edge of the system. The current vector field presented in 
Fig.~\ref{fig.current_c} is the sum of the contributions of the 
spin-$\uparrow$ and spin-$\downarrow$ current. 

Due to the breaking of time reversal symmetry, the spin-$\uparrow$ 
component dominates and thus resembles the ``quasi-helical'' situation 
described in Ref.~\cite{menard.guissart.17}. 
It is worth noting that a 
magnetic impurity or a ferromagnetic island proximity--coupled to a SOC 
superconductor will exhibit a finite spin polarization, thus giving 
rise to persistent currents 
\cite{pershoguba.bjornson.15,bjornson.pershoguba.15,li.neupert.16} as a 
result of the magnetoelectric effect. This type of bond current along the 
nanoflake can be observed experimentally, e.g. in the differential 
conductance \mbox{
measurements~\cite{drozdov.alexandradinata.14,palaciomorales.mascot.18}.}

\subsection{Numerical results for irregular nanoflake}
\label{sec.irregular}

In order to go beyond the circular 
limit discussed in previous paragraphs, we locate the substituted atoms at random sites of the lattice as nearest neighbors of an initial atom, located at the center of the surface.
This method results in a nanoflake with {\it rugged} boundary (cf. Fig.~\ref{fig.schemat}) between the substituted (blue) and substrate (gray) atoms, which is similar to experimental setups~\cite{menard.guissart.17,palaciomorales.mascot.18}.
However, artificial construction of nanoflakes could be characterized by a more regular shape too.
In practice, in the SM~\cite{Note1}, we show that the main properties of the system do not depend qualitatively on the shape of the nanoflake. Therefore, all properties described before remain when the flake becomes irregular.



\section{Proposal to experimentally measure the chirality using Scanning tunneling Microscopy }
\label{sec.exp_chiral}


\begin{figure}[!b]
\centering
\includegraphics[width=\linewidth]{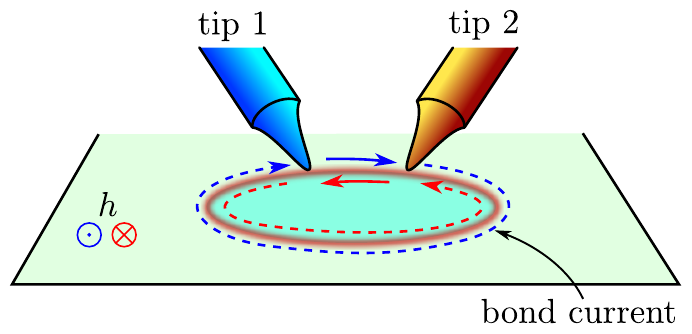}
\caption{
Setup of a double-tip experiment in order to probe the chirality of the edge states.
In the presence of the external magnetic field $h$ in $\uparrow$ $\odot$ (red) or $\downarrow$ $\otimes$ (blue) direction, the chiral bond current flows clockwise (blue arrows) or counterclockwise (red arrows) along the boundary of the system.
Thus, the double-tip measurement of non-local differential conductance $G_{12}$  depends on the chirality of the edge state, i.e. on the direction of the magnetic field and the bond current.
\label{fig.schmat_two_tip}
}
\end{figure}

Our experimental proposal is based on the double-tip measurement technique~\cite{shiraki.tanabe.01,kolmer.olszowski.17,voigtlander.cherepanov.18}.
The in-gap edge states localized around the nanoflake can give a {\it nonlocal} response between two spatially separated tips, which is schematically shown in Fig.~\ref{fig.schmat_two_tip}.
Similar transconductance technique was successfully used to measure an in-gap surface band~\cite{kolmer.brandimerte.19}.

We performed the calculation of the local and non-local differential conductance using the {\sc Kwant}~\cite{groth.wimmer.14} code to numerically obtain the scattering matrix~\cite{lesovik.sadovskyy.11,akhmerov.dahlhaus.11,fulga.hassler.11}:
\begin{eqnarray}
S = \left( \begin{array}{cc}
S_{11} & S_{12} \\ 
S_{21} & S_{22}
\end{array} \right) , \quad S_{ij} = \left( \begin{array}{cc}
S_{ij}^{ee} & S_{ij}^{eh} \\ 
S_{ij}^{he} & S_{ij}^{hh}
\end{array} \right) .
\end{eqnarray}
Here, $S_{ij}^{\alpha\beta}$ is the block of the scattering amplitudes of incident particles of type $\beta$ in tip $j$ to particles of type $\alpha$ in tip $i$.
Then, the differential conductance matrix is given as~\cite{rosdahl.vuik.18}:
\begin{eqnarray}
G_{ij} ( E ) \equiv \frac{\partial I_{i}}{\partial V_{j}} = \frac{e^{2}}{h} \left( T_{ij}^{ee} - T_{ij}^{he} - \delta_{ij} N_{i}^{e} \right) ,
\end{eqnarray}
where $I_{i}$ is the current entering terminal $i$ from the scattering region and $V_{j}$ is the voltage applied to the terminal $j$, and $N_{i}^{e}$ is the number of electron modes at energy $E$ in terminal $i$.
Finally, the energy transmission is:
\begin{eqnarray}
T_{ij}^{\alpha\beta} = \Tr \left( \left[ S_{ij}^{\alpha\beta} \right]^{\dagger} S_{ij}^{\alpha\beta} \right) .
\end{eqnarray}

The nonlocal response is constituted by two processes:\\
(i) a
 direct electron transfer between the leads and (ii) the crossed Andreev reflection (CAR) of an electron from one tip into a hole in the second tip~\cite{byers.flatte.95,deytscher.feinberg.00}. 
In typical cases, the CAR contribution dominates the electron transfer~\cite{falci.hekking.01,backmann.weber.04} and such processes is responsible for the  Cooper pair splitter~\cite{herrmann.portier.10}.
Here, we show that this technique has a potential application in  measuring the chirality of the edge state.
In order to achieve this goal, we suggest the use of two ferromagnetic tips described by the Hamiltonian:
\begin{eqnarray}
\mathcal{H}_{tip}^{i} = \sum_{{\bm k}\sigma} \left( \varepsilon_{{\bm k}\sigma} - \sigma M_{i} \right) c_{{\bm k}\sigma}^{\dagger} c_{{\bm k}\sigma} ,
\end{eqnarray}
where $\varepsilon_{{\bm k}\sigma}$ is the dispersion relation of the free electrons in the tips, while $M_{i}$ is the magnetization of the $i^{\rm th}$ tip.
We assume that the tips have opposite magnetization, i.e. $M_{1} = - M_{2}$.
The tips are separated from the plane of the system by the barrier potential.
We assume a system like that shown schematically in Fig.~\ref{fig.schmat_two_tip}, i.e. tips are located exactly above the nanoflake edge, in a non-symmetric position.
As a result, a distance between 1st and 2nd tip differs when measured clockwise or counterclockwise (along the edge state channel at the border of the nanoflake).

\begin{figure}[!t]
\centering
\includegraphics[width=\linewidth]{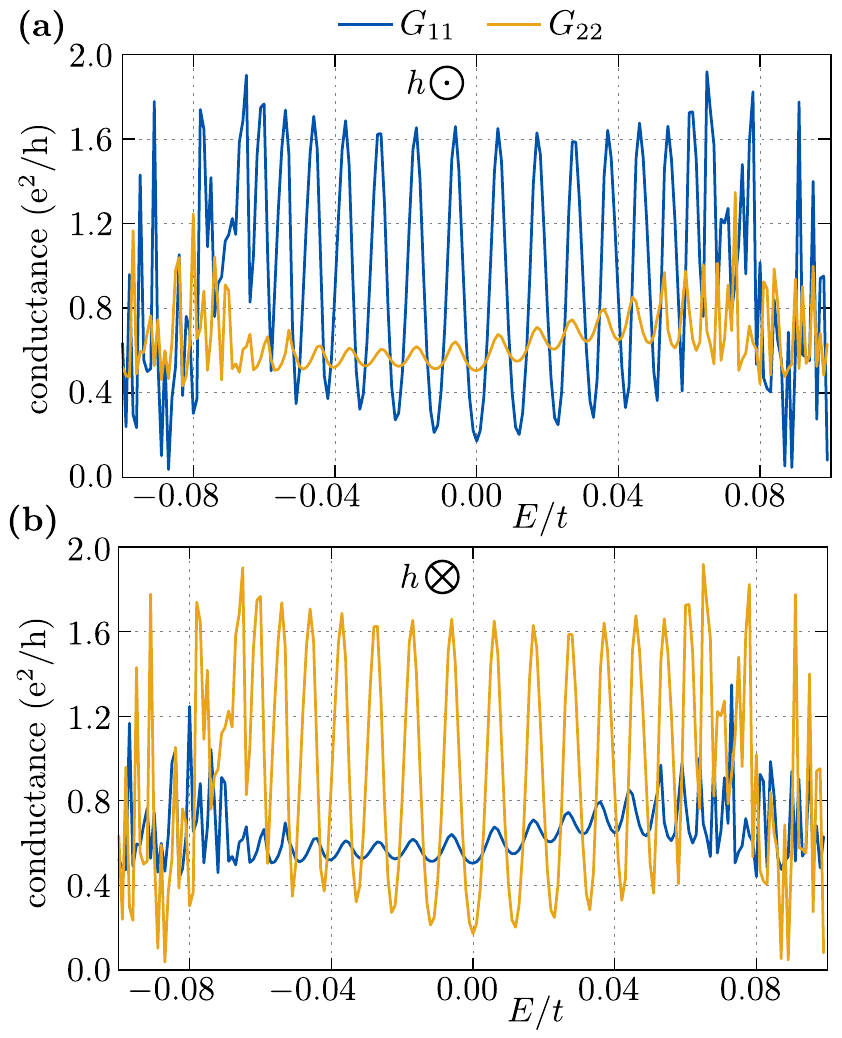}
\caption{
The local conductance $G_{11}$ and $G_{22}$ are shown in blue and red lines for positive (a) and negative (b) magnetic field $h/t= \pm 0.4$.
Here, we take two ferromagnetic (FM) tips with $M_{1} = -M_{2} = 3.0t$.
The direction of the magnetic field applied to the system is shown on top.
\label{fig.localg}
}
\end{figure}

\begin{figure}[!t]
\centering
\includegraphics[width=\linewidth]{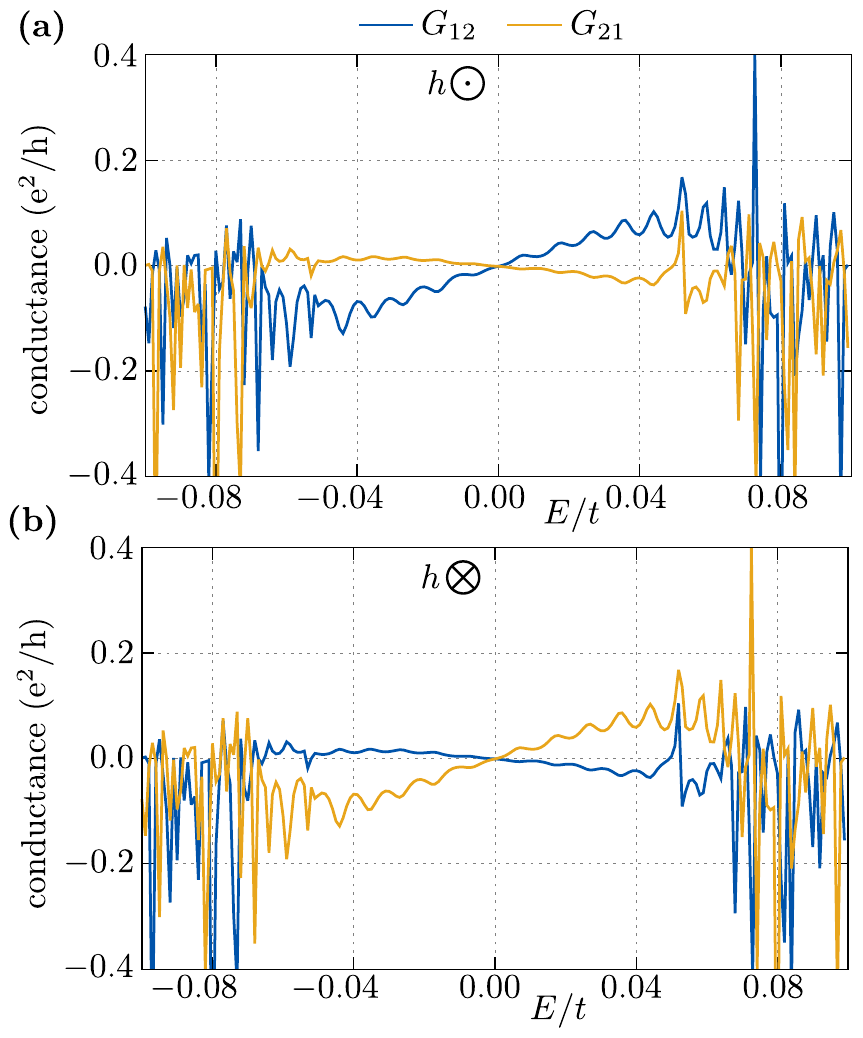}
\caption{
The same as Fig.~\ref{fig.localg}, but in the case of the non-local conductance $G_{12}$ and $G_{21}$, shown by red and blue lines, respectively.
The direction of the magnetic field applied to the system is shown on top.
\label{fig.nonlocalg}
}
\end{figure}

In our system, we can find the local ($G_{11}$ and $G_{22}$) as well as non-local ($G_{12}$ and $G_{21}$) differential conductance (Fig.~\ref{fig.localg} and Fig.~\ref{fig.nonlocalg}, respectively).
The local conductance $G_{ii}$ can be treated as a probe of the existence of states in the system~\cite{chene.yu.17,zhang.liu.18}. 
In this sense, each state gives a positive signal in $G_{ii}$, independently of the direction of the magnetic field $h$ (see top and bottom panels in Fig.~\ref{fig.localg}).
As we can see, in our case we observe several in-gap states.
Due to the use of ferromagnetic tips we observe non-equality of $G_{11}$ and $G_{22}$ (blue and red lines, respectively).

\begin{figure*}
\centering
\includegraphics[width=\linewidth]{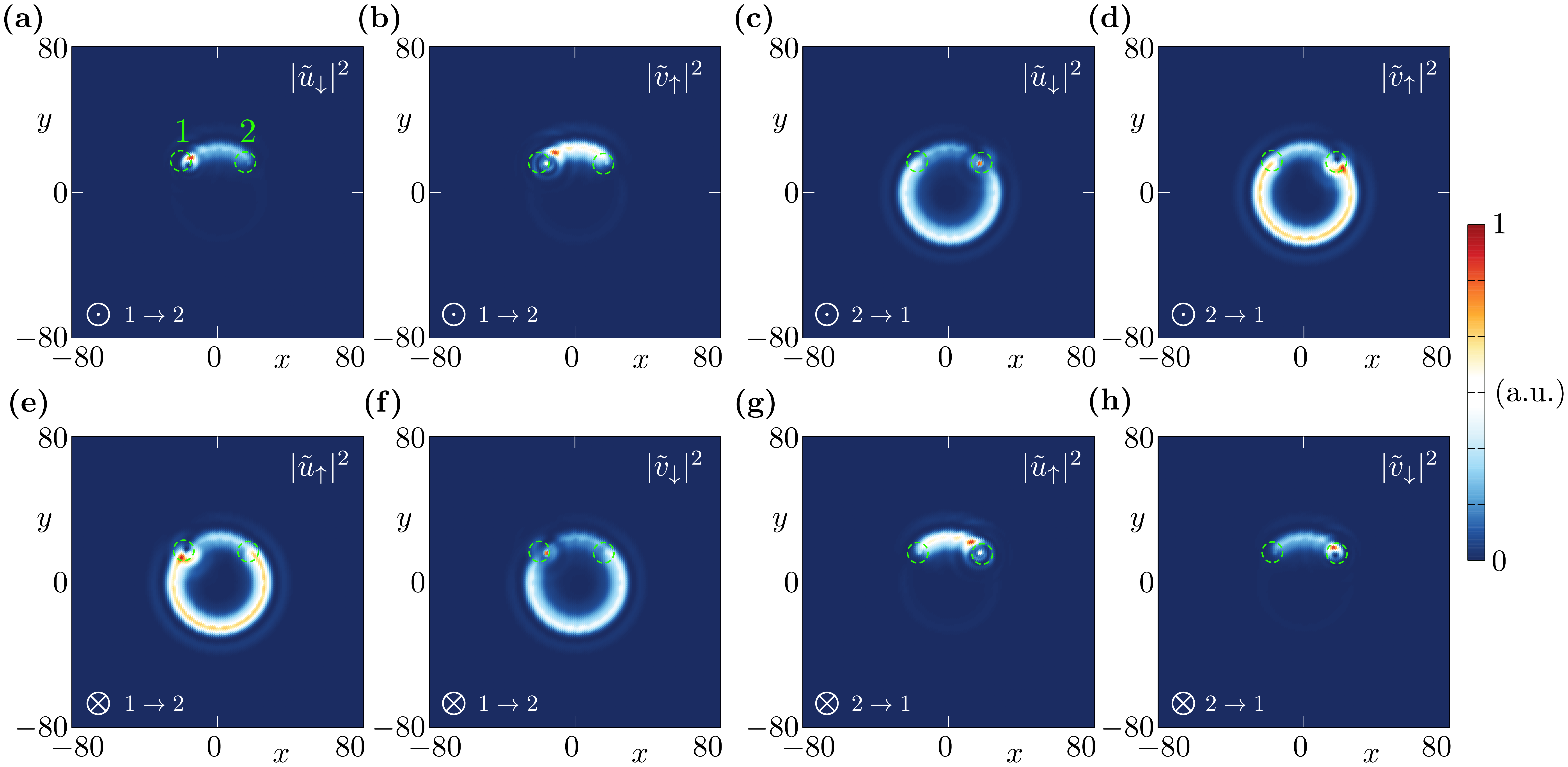}
\caption{
Spatial profile of the wavefunctions (particle- and hole-like component, given by $u$ and $v$ respectively) having the largest contribution to propagating modes at energy $0.02t$.
Top (bottom) panels show results in the case of positive (negative) value of magnetic field and are depicted by  $\odot$ ($\otimes$).
First two columns from left, correspond to propagating modes from 1st to 2nd tip ($1 \rightarrow 2$), while two column from right -- from 2nd to 1st tip ($2 \rightarrow 1$).
STM tips are labeled at panel (a).
Yellow dashed line represents the region of the system bordered by the potential barrier.
\label{fig.modes}
}
\end{figure*} 

The non-local conductances $G_{12}$ and $G_{21}$ (Fig.~\ref{fig.nonlocalg}) describe  different situations.
In order for CAR processes to occur, electrons from both tips need to have opposite spins to create a Cooper pair and simultaneously eject a hole  from the other tip.
Due to the applied external magnetic field the incident electrons from the edge bond current would have the same spin as the electrons from the first tip they encounter. 
The electron with opposite spin  should come the other tip (which has opposite magnetization) and emit a hole with the same spin  to constitute a CAR process.
Thanks to this nonlocal phenomena, we can easily determine the direction of the edge state propagation as the nonlocal conductance coincides with the direction of chiral bond current.
Thus, if the sign of magnetic field  changes,  the direction of the edge state propagation reverts too, $G_{12}$ becomes $G_{21}$ due to the spatial symmetry of the system.
However, tips are not perfectly magnetized, therefore nonlocal conductance measurement in the direction opposite to the chiral bond current remains nonzero (\textit{e.g.} $G_{21}$ for $\odot$ direction of magnetic field).
As the edge state bond current is constituted of both spins, particles with spins not aligned with the tip magnetization scatter off the tip (\textit{cf.} wavefunction localizations in Fig.~\ref{fig.modes}).
If the magnetic field is not present in the system, in-gap nonlocal conductance vanishes 
(not shown), and peaks at the edge of superconducting gap appear~\cite{rosdahl.vuik.18}.
Differences between the absolute value of non-local conductances are the consequence of the non-symmetric position of the tips.
The change from negative to positive slope near the gap can be interpreted as a crossover from subgap transport dominated by crossed Andreev reflection to charge imbalance above the gap~\cite{kalenkov.zaikin.11}.
Additionally, the non-local conductance strongly depends on a distance between the tips~\cite{rosdahl.vuik.18}.

To explain the above results, we analyzed the propagating modes in the system (Fig.~\ref{fig.modes}).
The nonlocal transport corresponds to situation where incident electron from one tip is transmitted through the edge modes as the chiral mode and is scattered into the second tip. 
Such propagating modes can be represented in form of wavefunction $\psi({\bm r}) = ( \tilde{u}_{\downarrow} , \tilde{u}_{\uparrow} , \tilde{v}_{\downarrow} , \tilde{v}_{\uparrow} )^{T}$, where $\tilde{u}$ and $\tilde{v}$ correspond to its electron- and hole-like component, respectively. 
Here,  $\psi$ is composed mostly of $\downarrow$-electron and $\uparrow$-hole component~\cite{ikeagaya.asano.19}.
Regions with the non--zero  probability of localization correspond to the particle remnants of scattering processes, which distribution coincides with the direction of edge state propagation.
As we mentioned above, with relatively small voltage bias, the non-local transport is dominated by CAR processes (which was envisioned by the antiparallel magnetization of $G_{12}$ and $G_{21}$).
In the case of the ``positive'' $\odot$ magnetic field, we observe propagation from 1st to 2nd tip in clockwise direction [Fig.~\ref{fig.modes}, panels (a) and (b)], for both electron and hole  component of the wavefunction.
Then, if we check the mode propagation from 2nd to 1st tip [Fig.~\ref{fig.modes}, panels (c) and (d)], we can see that the chirality of the propagating modes is preserved and clockwise.
In the case of the magnetic field with opposite direction, we observe modes propagating in opposite direction as previously mentioned (bottom panels in Fig.~\ref{fig.modes}).
Without any surprise, mode propagation direction in this case is also preserved.

Summarizing, the non-local conductance is not just a fingerprint of the existence  of the chiral mode~\cite{ikeagaya.asano.19}, but also a tool to measure its chirality.



\section{Summary and conclusions}
\label{sec.sum}

Recent experimental results have presented the possibility of the emergence of non--trivial topological phases in magnetic nanostructures coupled to superconducting substrates~\cite{menard.guissart.17,palaciomorales.mascot.18}.

In this paper, we have explored the artificial implementation of topological phase transitions induced by the local modification of the chemical potential.
In this respect, we performed analytic calculations valid in the continuum limit in the case of a nanoflake with a circular geometry and found the spectrum of the system as a function of the total angular momentum. We have also studied how the transverse spatial extent of the wave function of the chiral Majorana state localized around the nanoflake depends upon the system parameters. 
We then performed similar calculations for a finite size geometry using a  tight binding formulation.
In-gap states correspond to prominent peaks in the LDOS only in a distinct region of space, identified as the domain wall, which should be observed relatively simply through STM experiments. 
We have introduced a real space indicator, which locally characterizes  the topological phase. 
Indeed, for a few sets of parameters, results obtained from this indicator were in agreement with those obtained in the continuum limit.
In the case analyzed here, the effective magnetic field leads to the realization of a 
non--trivial phase only in distinct regions of the system, creating a non--trivial superconducting dome surrounded by a trivial superconducting phase. 
This phase separation could be observed through the measurement of a bond chiral current, which is connected to the existence of strongly localized in-gap states.
We have shown that this current circulates around the nanoflake.
Additionally, with the help of the indicator, we have found the topological phase diagram of the system. 
We have shown that an artificial phase separation can be induced for a finite range of effective magnetic field. 
The boundary of this phase separation is strongly related to the effective magnetic field, which determines the transition between
trivial to non--trivial phases

In the last part of our work, we have proposed an experimental method to measure the chiraliy of the edge states, based on a double-tip measurement of the non-local differential conductivity. We have found that
the non-local transport properties between the two tips allow one to determine  the chirality of the edge state. Although challenging, this type of experiment should be accessible with the present technology.

\begin{acknowledgments}
We would like to thank Tristan Cren for interesting discussions.
This work was supported by the National Science Centre (NCN, Poland) 
under grants Nos.:
2017/24/C/ST3/00276 (A.P.),   
2017/27/N/ST3/01762 (S.G.),   
2018/31/N/ST3/01746 (A.K.),   
2016/23/B/ST3/00839 (A.M.O.), 
and
2017/25/B/ST3/02586 (P.P.).   
A.~M.~Ole\'s is grateful for the Alexander von Humboldt
Foundation Fellowship \mbox{(Humboldt-Forschungspreis)}.
AP appreciate also founding in the frame of scholarships of the Minister of Science and Higher Education (Poland) for outstanding young scientists (2019 edition, no. 818/STYP/14/2019).
\end{acknowledgments}

\bibliography{biblio}

\end{document}